\documentstyle[multicol,aps,psfig]{revtex}
\begin{document}
\newcommand{\la} {\langle}
\newcommand{\ra} {\rangle}
\newcommand{\ep} {\varepsilon}
\newcommand{\bu} {\bf u}
\newcommand{\de} {\delta}
\pagestyle{myheadings}
\draft
\preprint{Submitted to Physics of Fluids}
\title{Lattice-Boltzmann Simulations of Fluid Flows in MEMS}
\author{Xiaobo Nie$^{1}$, Gary D. Doolen$^{1}$ and Shiyi Chen$^{1,2}$ }
\address{${}^{1}$Center for Nonlinear Studies and Theoretical Division,
Los Alamos National Laboratory, Los Alamos, NM 87545\\
${}^{2}$IBM Research Division, T. J. Watson Research Center,
P.O. Box 218, Yorktown Heights, NY 10598\\}

\maketitle

\begin{abstract}
The lattice Boltzmann model is a simplified kinetic method
based on the particle distribution function. We 
use this method to simulate problems in MEMS, in which 
 the velocity slip near the wall plays an important role. 
It is demonstrated that the lattice Boltzmann method 
can capture the fundamental behavior in micro-channel flow, including
velocity slip and nonlinear pressure drop
along the channel.  The Knudsen number dependence of the position of the vortex 
center and the pressure contour in micro-cavity flows is also demonstrated. 
\end{abstract}

\begin{multicols}{2}
\narrowtext

	The development of technologies in Micro-electro-mechanical 
systems (MEMS) has motivated the study of fluid flows in devices 
with micro-scale geometries, 
such as micro-channel and micro-cavity flows \cite{Ho}. In these flows, 
the molecular mean free path of fluid molecules could be the same order 
as the typical geometric length of the device;  then the 
continuum hypothesis which is the fundamental for the
Navier-Stokes equation breaks down. 
An important feature in these flows is the emergence of a slip velocity
at the flow boundary, which strongly affects the 
mass and heat transfer in the system.
In micro-channel experiments, it has been observed that the measured mass 
flow rate is higher than that based on a non-slip boundary condition 
\cite{Arkilic}. The Knudsen number, $K_n= {l}/{L}$,
can be used to identify the influence of the effects of the mean free 
path on these flows, where $l$ is the mean free 
path of molecules and $L$ is the typical length of the flow domain. 
It has been pointed out that for a system with $K_n<0.001$, the fluid flow
can be treated as continuum. For $K_n>10$ the system 
can be considered as a free-molecular flow. 
The fluid flow for $0.001<K_n<10$, which often appears in 
the MEMS \cite{Ho}, 
can not be considered as a continuum nor a free-molecular flow.
Traditional kinetic methods, such as molecular dynamics 
simulations\cite{Koplik} and the 
continuum Boltzmann equation approach, could be used to describe 
these flows. But these methods are more 
complicated than schemes usually used for continuum hydrodynamic 
equations. The solution of the Navier-Stokes equation including  
the velocity-slip boundary condition with a variable parameter   
has also been used to simulate micro-channel flows \cite{Beskok}.

In the past ten years, the lattice Boltzmann method
 (LBM)\cite{chen-doolen} has emerged as an 
alternative numerical technique for simulating fluid 
flows. This method solves a simplified Boltzmann equation on 
a discretized lattice.  
The solution of the lattice Boltzmann equation
 converges to the Navier-Stokes solution 
in the continuum limit (small Knudsen number). In addition,
since the lattice Boltzmann method is intrinsically 
kinetic, it can be also used to simulate fluid flows with
high Knudsen numbers, including fluid flows in very small MEMS. 

To demonstrate the utility of the LBM, we use the LBM model with
three speeds and nine velocities on a two-dimensional square
lattice. The velocities, ${\bf c_i}$, include eight moving velocities
along the links of the square lattice and a zero velocity for the rest
particle. They are: $(\pm1,0),(0,\pm1),(\pm1,\pm1), (0,0)$. 
Let $f_i({\bf x},t)$ be the distribution functions at  
${\bf x}$, $t$ with velocity ${\bf c_i}$. 
The lattice Boltzmann equation with the BGK collision  
approximation\cite{Chen91,Qian92} can be written as
\begin{equation}
 f_i({\bf x}+{\bf c_i}\delta x ,t+\delta t)-f_i({\bf x},t)= 
-{\tau}^{-1} (f_i-f^{eq}_i),
\label{kin}
\end{equation}
where $f^{eq}_i (i=0,1,\cdot\cdot\cdot,8)$ 
is the equilibrium distribution 
function and $\tau$ is the relaxation time. We have assumed that
the spatial separation of the lattice is $\delta x$ and the time step
is $\delta t$.   A suitable equilibrium 
distribution is\cite{Qian92}:
\begin{eqnarray}
f_i^{eq} = t_i \rho \left [ 1 + \frac{c_{i\alpha} u_{\alpha}}{c_s^2}
+\frac{(c_{i\alpha}c_{i\beta}-c_s^2\delta_{\alpha\beta})}{2c_s^4}
u_{\alpha}u_{\beta} \frac{}{} \right ].
\label{equil}
\end{eqnarray}
Here $c_s = 1/\sqrt{3}$, $t_0 = 4/9, t_1 = t_2 = t_3 = t_4 = 1/9$ and 
$t_5 = t_6 = t_7 = t_8 = 1/36$. 
The Greek subscripts $\alpha$ and $\beta$ denote the spatial directions
in Cartesian coordinates. The density $\rho$ and the fluid velocity ${\bf v}$ 
are defined by 
$\rho=\sum_i f_i$, $\rho {\bf v}=\sum_i {\bf c_i} f_i $. In previous
lattice-BGK models, $\tau$ was chosen to be a constant. This is applicable 
only for nearly-incompressible fluids. In micro-flows,
the local density variation is still relatively small, but
the total density change, for instance the density 
difference between the inlet and exit of a very long channel, could be
quite large. To include the dependence of viscosity on density 
we replace $\tau$ in Eq.(\ref{kin}) by $\tau'$: 
$
\tau'=\frac{1}{2}+\frac 1 {\rho} (\tau-\frac 1 2).
$
Using the Chapman-Enskog multi-scale expansion technique, 
we obtain the following Navier-Stokes 
equations in the limit of long wavelength and low Mach number:
\begin{eqnarray}
\label{density}
&\partial_t \rho + \partial_{\alpha} (\rho u_{\alpha})  =  0, \\
\label{mom}
&\partial_t (\rho u_{\alpha}) + \partial_{\beta} (\rho u_{\alpha} u_{\beta})
 =  -\partial_{\alpha} P -\partial_{\beta}\pi_{\alpha \beta}, \\
&P  =  c_s^2\rho, \ \ \pi_{\alpha \beta}  =
\nu (\partial_{\alpha} (\rho u_{\beta}) + 
\partial_{\beta} (\rho u_{\alpha})), \nonumber
\end{eqnarray}
where $\nu = {c_s^2}(2\tau-1)/({2 \rho})$ is the kinematic viscosity. 
In classical kinetic theory, the viscosity $\nu$ for
a hard sphere gas is linearly proportional to the mean free path. Similarly,
we define the mean free path $l$ in the LBM as: $a (\tau - 0.5)/\rho$, 
where $a$ is constant. 

	Our first numerical example is a micro-channel flow \cite{Arkilic}.
The flow is contained between two parallel plates separated by a distance 
$H$ and driven by the pressure difference between the inlet pressure, $P_i$,
 and exit pressure, $P_e$.
The channel length in the longitudinal direction is $L$. 
We take $L=1000, H=10$ (lattice units) in our simulations satisfying 
$L/H>>1.$ 
The bounce-back boundary condition is used 
for the particle distribution functions at the top and bottom plates, i.e.,
when a particle distribution hits
a wall node, the particle distribution scatters back to the fluid node
opposite to its incoming direction. 
A pressure boundary condition is used at the input and the exit.
\vskip -10pt
\begin{figure}
\centerline{\psfig{file=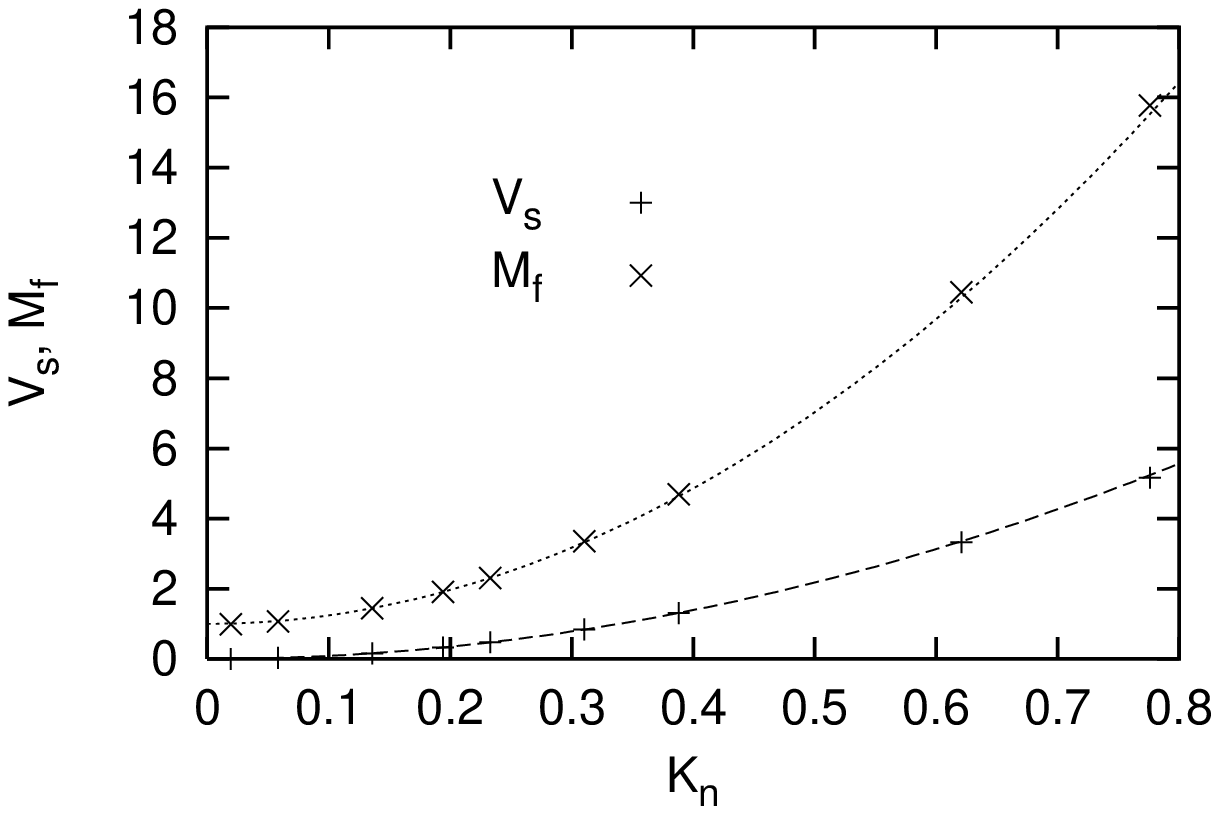,width=220pt}}
{\small FIG.~1. The slip velocity and the normalized mass flow rate
at the exit of a micro-channel flow as functions of $K_n$
for $P_i/P_e = 2$. The `+' and `$\times$' are LBM numerical results.
The dashed and dotted lines are Eq.(\ref{slip}) and Eq.(\ref{mf}) respectively. }
\label{fig1}
\end{figure}
The slip velocity $V_s$ at the exit of the micro-channel flow
is defined as: 
$
u(y)=u_0(Y-Y^2+V_s),
$
where $u(y)$ is the velocity along the $x$ (or flow) 
direction at the exit and $Y=y/H$. $u_0$ and $V_s $ can be obtained by 
fitting numerical results using the least squares method. This definition
of the slip velocity is consistent with others\cite{Arkilic,Beskok}.
In Fig.~1, we plot the slip velocity $V_s$ and the 
normalized mass flow rate $M_f = M/M_0$, 
as functions of Knudsen number when the pressure ratio 
$\eta=P_i/P_e = 2$. The normalization factor,
$M_0=\frac{h^3 P_e}{24 \nu} (\eta - 1)$, is the mass flow rate when the 
velocity slip is zero. To calculate the Knudsen number we have chosen 
$a=0.388$ in order to match the simulated mass flow rate  
with experiments (See theory curve in Fig.~3). 
Using a least squares fit to the data in Fig.~1, we obtain: 
\begin{eqnarray}
V_s=8.7 K_n^2. 
\label{slip}
\end{eqnarray}
If we assume that the Navier-Stokes equations are 
valid for the micro-flows except that the slip boundary condition 
$V_s$ in Eq.(\ref{slip}) replaces the traditional
non-slip condition on the walls \cite{Beskok}, Eqs. (\ref{density}), (\ref{mom}) 
and (\ref{slip}) will give the mass flow rate:
\begin{eqnarray}
 M_f=1+12V_s(K_n) \frac{\ln (\eta)}{\eta^2-1}.
\label{mf}
\end{eqnarray} 
For $\eta=2$, the above formula becomes $M_f=1+24.1K_n^2$, 
which agrees well with the numerical results in Fig~1.

In laminar Poiseuille flows, one usually assumes that the density 
variation along the channel is very small, and the pressure drop 
along the channel is nearly linear. 
In micro-channel flow, however, the ratio between the length and the 
width is much larger and the pressure drop is not linear. 
If there is no velocity slip at the walls, it has been shown\cite{Arkilic,Beskok} 
from the Navier-Stokes equation that the pressure along the channel 
has the following dependence on the dimensionless coordinate, $X=x/L$:
\begin{eqnarray}
P^2=P_e^2[1+(\eta^2-1)(1-X)],
\label{p1}
\end{eqnarray}
If the velocity at the boundaries is allowed to slip, the pressure drop along 
the channel will depend on the Knudsen number. 
In Fig.~2 we present the LBM simulation results for the normalized
pressure deviation from a linear pressure drop, 
$(P-P_l)/P_e$, as functions of $X$ for several Knudsen 
numbers, where $P_l=P_e + (P_i-P_e)(1-X)$. 
It is seen that when $K_n \le 0.2$, $(P-P_l)/P_e$ is a positive
 nonlinear function of $X$. This 
agrees with the results in \cite{Beskok} using an engineering model.
For $K_n \ge 0.2$, the LBM simulation shows that $(P-P_l)/P_e$
becomes negative, which is directly linked to the fact that
the slip velocity depends on the square of $K_n$ in the LBM. 
For large $K_n$, the pressure can be derived from Eq.(\ref{mf}): 
\begin{eqnarray}
P =P_e[ \eta^{(1-X)}].
\label{p2}
\end{eqnarray}
The negative deviation from a linear pressure drop has not been experimentally 
observed before and it would be interesting to testify this experimentally. 
\begin{figure}
\centerline{\psfig{file=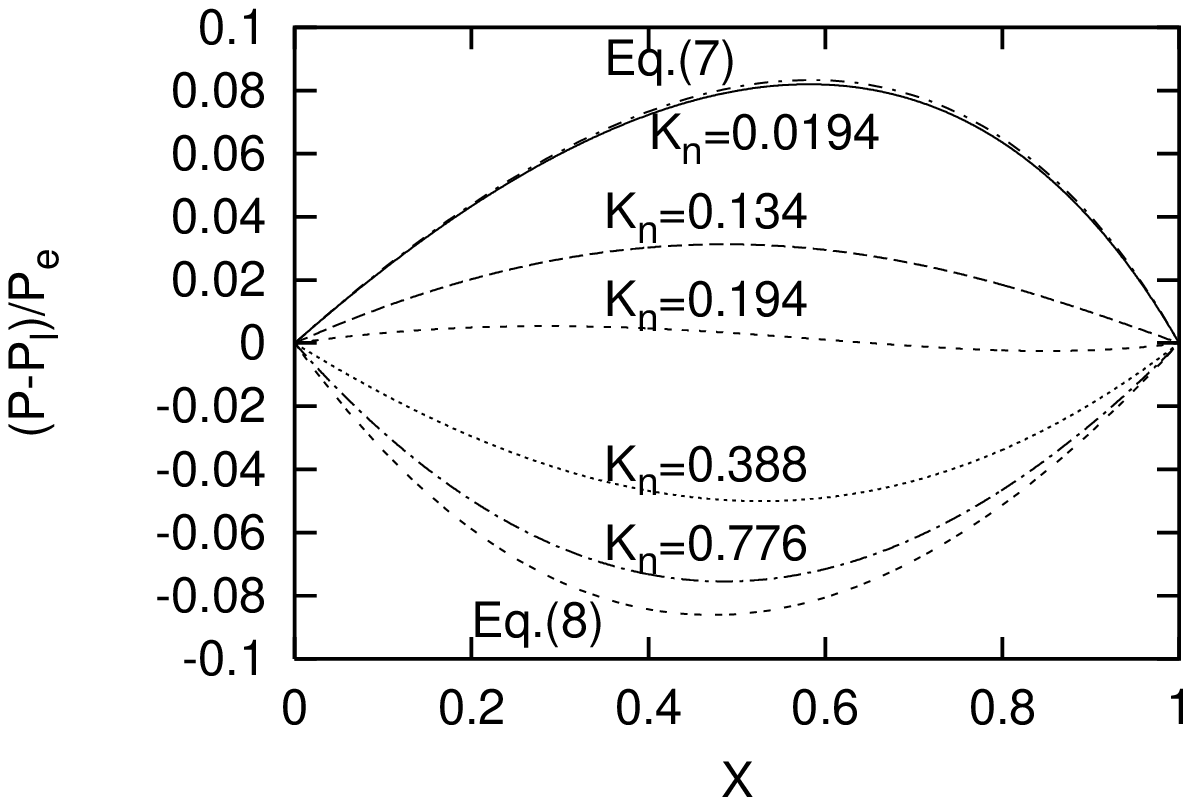,width=220pt}}
{\small FIG.~2. The deviations from linear pressure drop
for $\eta=P_i/P_e = 2$. The top and bottom lines are the analytical results
from Eq.(\ref{p1}) for $K_n=0$ and Eq.(\ref{p2}) for $K_n >> 1$ respectively.
The other curves are LBM numerical results for the Knudsen
numbers indicated. }
\label{fig2}\end{figure}

In Fig.~3 the mass flow rates as functions of the pressure ratio $\eta$ 
when $K_n=0.165$ are shown for our theory,  
the experimental work \cite{Arkilic},
the engineering model\cite{Beskok} and the LBM simulation.
Our theory and the LBM simulation agree well with the experimental 
measurements. It is noted that for large pressure ratios 
($\eta \ge 1.8$), the LBM agrees reasonably well with Beskok 
et al.\cite{Beskok}.  But for smaller pressure ratios, the difference 
increases because of different dependence of the slip velocity 
on $K_n$. 
\begin{figure}
\centerline{\psfig{file=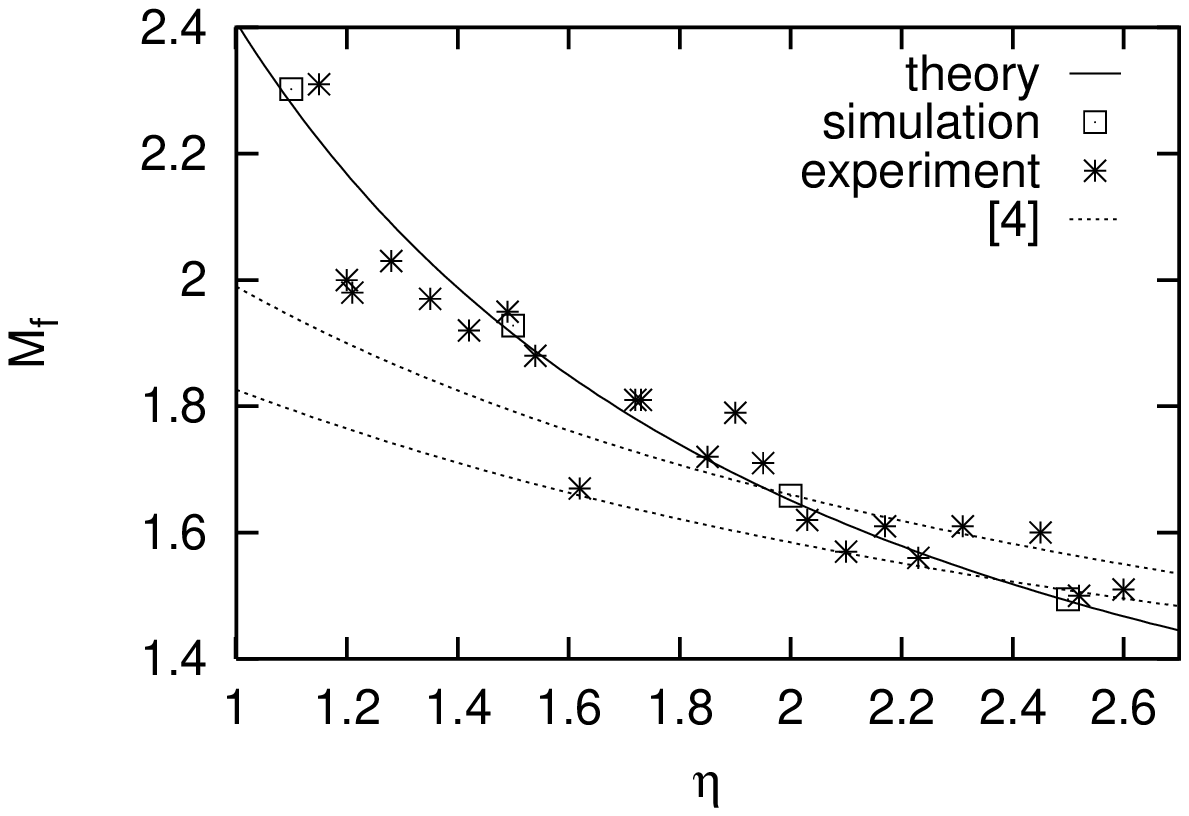,width=220pt}}
\vskip -10pt
{\small FIG.~3. The normalized mass flow rate as a function of the
pressure ratio for $K_n = 0.165$. The theory is Eq.(\ref{mf}). }
\label{fig3}
\end{figure}
%
\begin{figure}
\centerline{\psfig{file=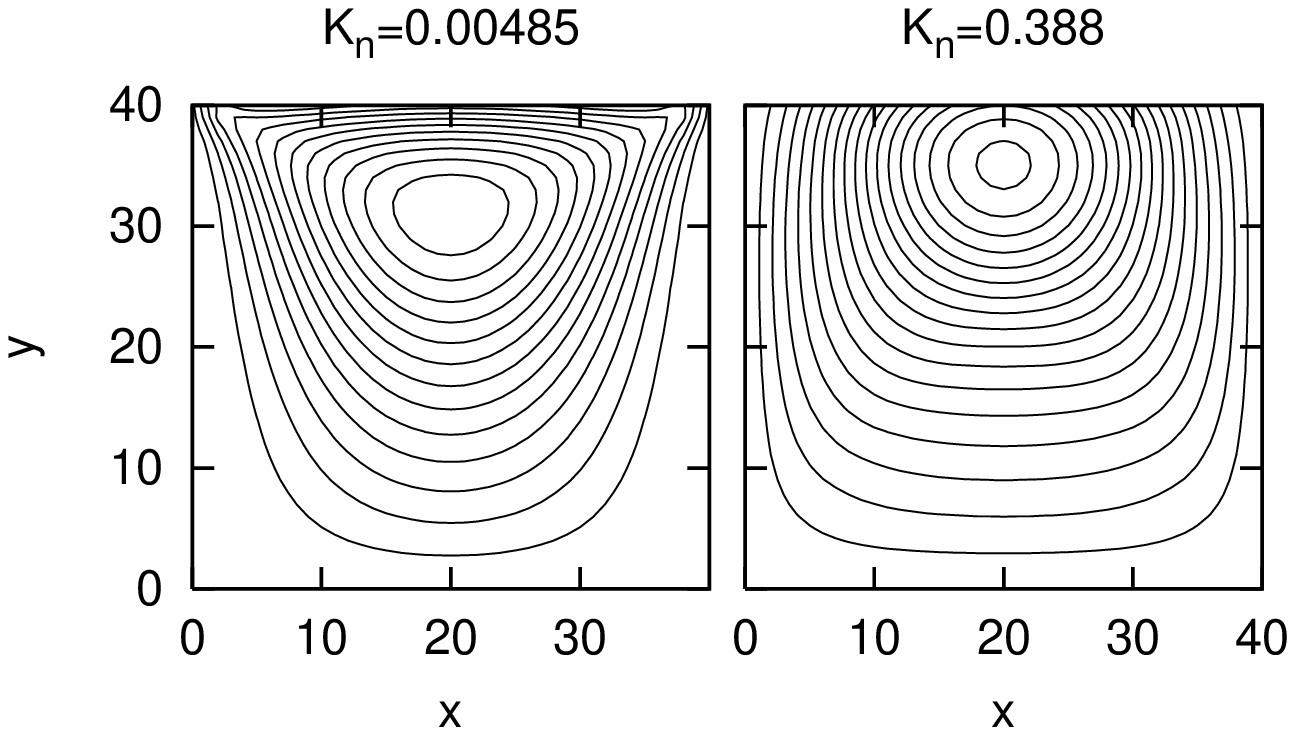,width=220pt}}
\vskip -10pt
{\small FIG.~4. Streamlines for two Knudsen numbers. }
\label{fig4}
\end{figure}
\vskip -10pt
\begin{figure}
\centerline{\psfig{file=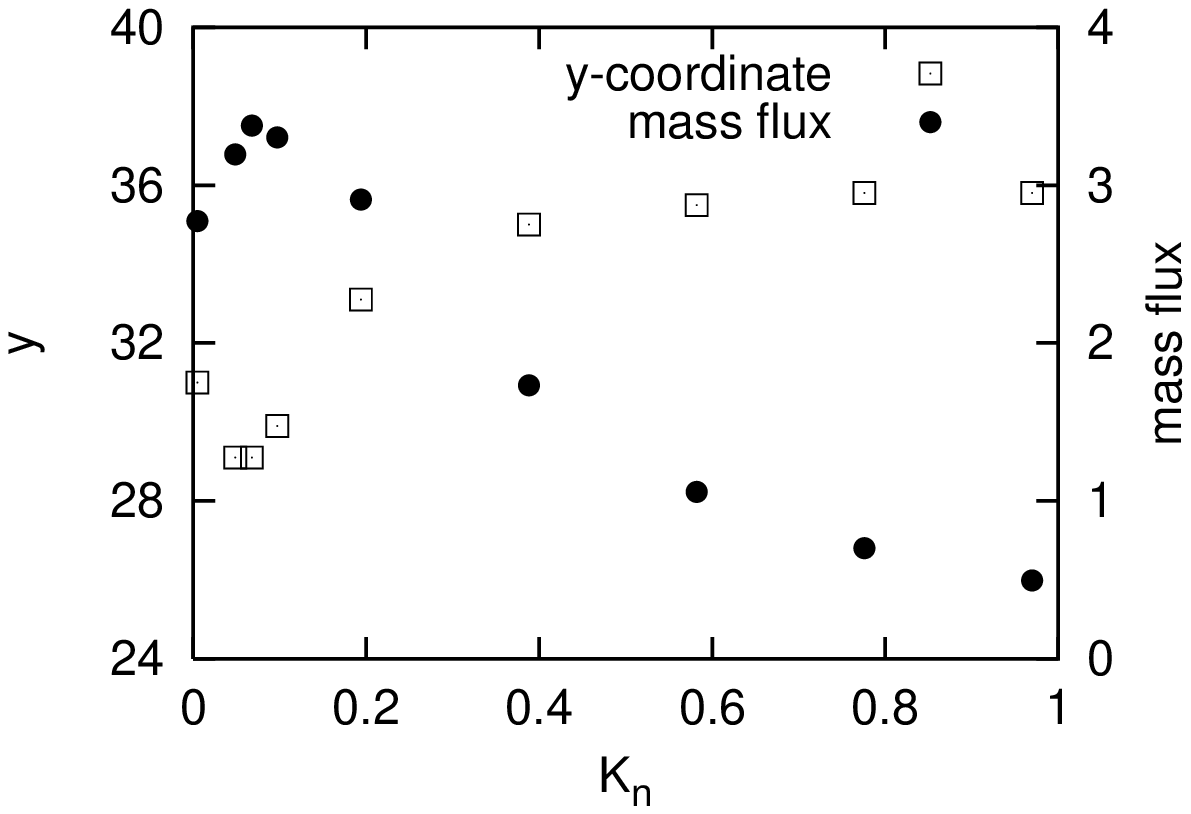,width=220pt}}
\vskip -10pt
{\small FIG.~5. The $y$-coordinate of the vortex center (square symbols) and
the mass flow(solid circles) as a function of $K_n$.
}
\label{fig5}
\end{figure}
%

Our second LBM numerical simulation is the two-dimensional micro-cavity flow. 
The cavity size is $L_x = L_y = 40$ (lattice units). The upper wall 
moves with a constant velocity, $v_0$, from left to right.  The other 
three walls are at rest, and bounce-back boundary conditions
are used. To see the dependence on  the Knudsen number in our simulations, 
we fixed the Reynolds number, $R_e = \frac{v_0 L_x}{\nu} = 2.4 \times 10^{-4}$
 and require the Mach number to be small, $M_a=\frac{v_0}{c_s} \le 10^{-3}$.
In Fig.~4, we show the streamlines for two different Knudsen numbers.
In Fig.~5 we show the vertical positions 
of the vortex center and the mass flux  
between the bottom and the vortex center as functions of $K_n$.
It can be seen that the center moves upward and the mass flow decreases 
with increasing Knudsen number. 
This occurs because the slip velocity on the upper wall 
causes momentum transfer to be less efficient. 
It has been shown\cite{Shuling} that the center of the vortex moves downward 
when the Reynolds number increases for very small $K_n$. 
Fig.~6 shows the pressure contours for the same parameters as in Fig.~4. 
Totally different pressure structures are observed for
these two cases. When the Knudsen number is small, the continuum assumption
is valid and the pressure contours are almost circles with centers 
at the left or the right corners. On the other hand, due to the
slip velocity on the walls, the pressure contours become nearly 
straight lines at the higher Knudsen number.
\begin{figure}
\centerline{\psfig{file=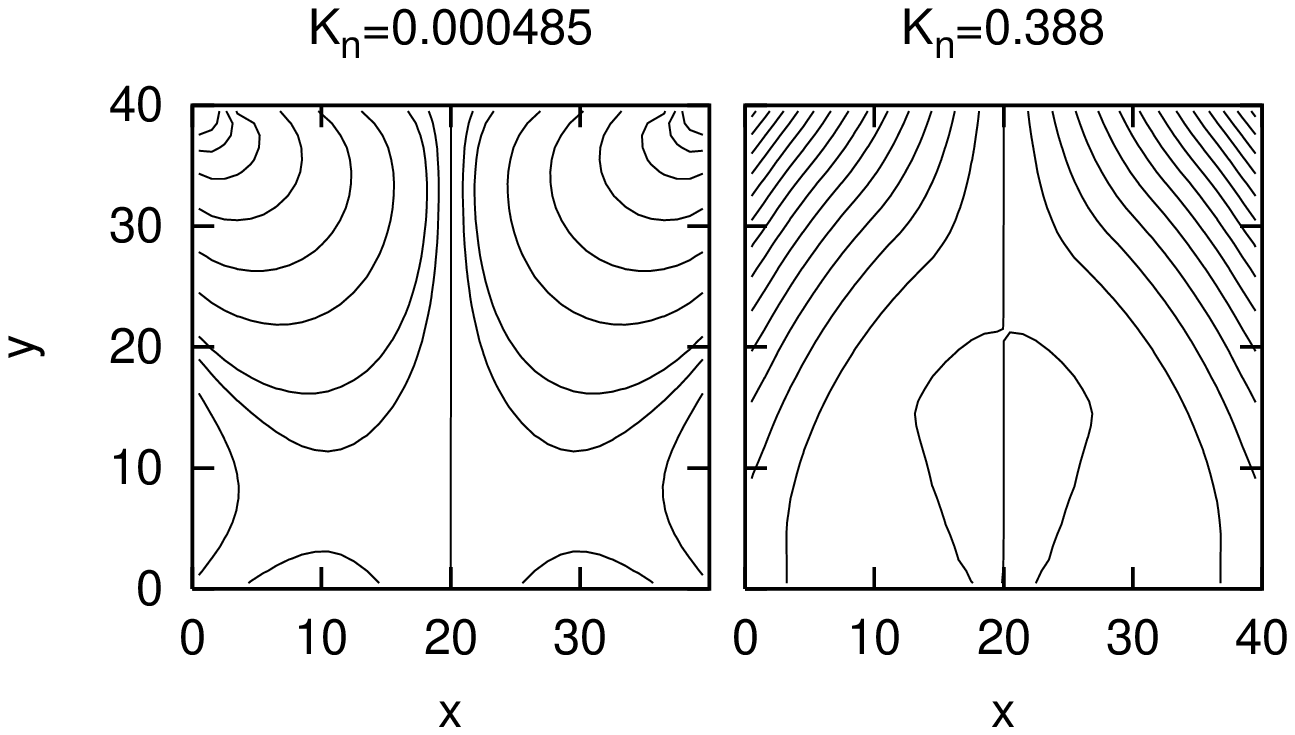,width=220pt}}
\vskip -10pt
{\small FIG.~6. The contours of pressure for the same two Knudsen numbers
shown in Fig.~4.  }
\label{fig6}
\end{figure}
\vskip -20pt

\end{multicols}
\end{document}